\documentclass[letterpaper, 10 pt, conference]{ieeeconf} \IEEEoverridecommandlockouts                             

\usepackage{amssymb}
\usepackage{makecell}
\usepackage{amsmath}
\usepackage{color,xcolor}
\usepackage{algorithm, algpseudocode}

\usepackage[utf8]{inputenc}
\usepackage{booktabs, caption, makecell}

\usepackage{threeparttable}
\usepackage{bbm}
\usepackage{tikz}
\usetikzlibrary{positioning}
\usepackage{multirow}
\usepackage[figuresright]{rotating}
\usepackage{amsmath}

\usepackage{tablefootnote}

\usepackage{tikz-network}
\usepackage{algorithm, algpseudocode}

\usepackage{multirow}
\usepackage{subcaption}

\title{\LARGE \bf Learning to Explain Solutions of Optimal Control Problems*}

\author{Jiyong Lee$^{1}$ and Ilias Mitrai$^{1,2}$
\thanks{*This work was supported by the McKetta Department of Chemical Engineering.}
\thanks{$^{1}$Jiyong Lee and Ilias Mitrai are with the McKetta Department of Chemical Engineering, The University of Texas at Austin, Austin, TX 78712, USA}
\thanks{$^{2}$Corresponding author. Email: {\tt\small imitrai@che.utexas.edu}}}

\usepackage{lipsum}

\begin{document}

\maketitle
\thispagestyle{empty}
\pagestyle{empty}

\begin{abstract}
In this paper, we propose an explainable artificial intelligence framework to explain the solution of optimal control problems predicted with graph neural networks (GNNs). The proposed approach first trains a GNN that predicts the solution of the optimal control problem for a given instance represented as a graph. Given the trained GNN, we use explainable AI algorithms to identify variables, constraints, variable--constraint connections, and parameters, i.e., nodes, edges, and features in the graph representation of the optimal control problem, that affect the prediction of the optimal solution the most. We apply the proposed approach to a case study regarding the optimal control of a continuous stirred tank reactor. First, the results show that the GNN model can accurately predict the optimal values of the manipulated variables. Application of explainable AI algorithms reveals equality and inequality constraints that are the most important for predicting the optimal solution. The inequality constraints correspond to ramping constraints, some of which are active at the optimal solution. 
\end{abstract}

\section{Introduction}

Constrained optimal control problems arise in many applications across science and engineering, such as process control and supply chain management. A characteristic example is model predictive control, where the control action is computed by repeatedly solving an optimization problem. Usually, the mapping between the problem formulation, i.e., variables, constraints, and parameters, and the optimal solution is treated as a black box. In general, a solver can provide a solution and a guarantee of local or global optimality, subject to numerical tolerances, but it does not explain why the specific solution is optimal. This can limit trust and implementation of the solution, especially when the returned solution is counterintuitive.

Traditionally, analysis of the optimal solution has relied on tools from optimization theory. For example, Lagrangian multipliers can quantify how perturbations in constraint coefficients affect the optimal objective value \cite{greenberg1993analyze}. Similarly, the set of active constraints provides information about the inequalities that bound the optimal solution. Recent approaches either focus on ``what-if" analysis using counterfactual explanations \cite{kurtz2026counterfactual} in tandem with inverse optimization \cite{korikov2021counterfactual, korikov2023objective} and large language models \cite{chen2024diagnosing}, aim to identify a subset of the variables and constraints that affect the solution \cite{rathi2024enhancing}, or compare the solution with previously obtained ones \cite{aigner2024framework}. Despite these advances, application of such approaches to explaining the solution of optimal control problems, especially in online settings, is challenging. For example, some approaches are applicable only to certain classes of optimization problems, e.g., linear programming problems, or require the repeated solution of the optimization problem under consideration for different parameter values, which can incur additional computational cost. 

In parallel, machine learning models have been used extensively to predict solutions to optimal control problems. These approaches approximate the mapping between problem parameters, e.g., set-point, and the optimal solution using a vectorial feature representation of the problem. For example, the features can capture the set-point and initial conditions of an MPC problem, and the label can be the first control action to implement \cite{kumar2021industrial, hertneck2018learning, karg2018deep, cauligi2022prism, mitrai2026discovering} or the entire trajectory \cite{11615398}. Although these approaches can have good predictive accuracy, they are black-box. Explainable artificial intelligence tools \cite{ribeiro2016should, lundberg2017unified} can quantify feature importance. However, application to ML models that use a vectorial representation of the optimal control problem cannot provide insights about the role of the constraints since constraint-relevant information is not typically provided as an input to the ML model. 

\begin{figure}
    \centering
    \includegraphics[width=0.8\columnwidth]
    {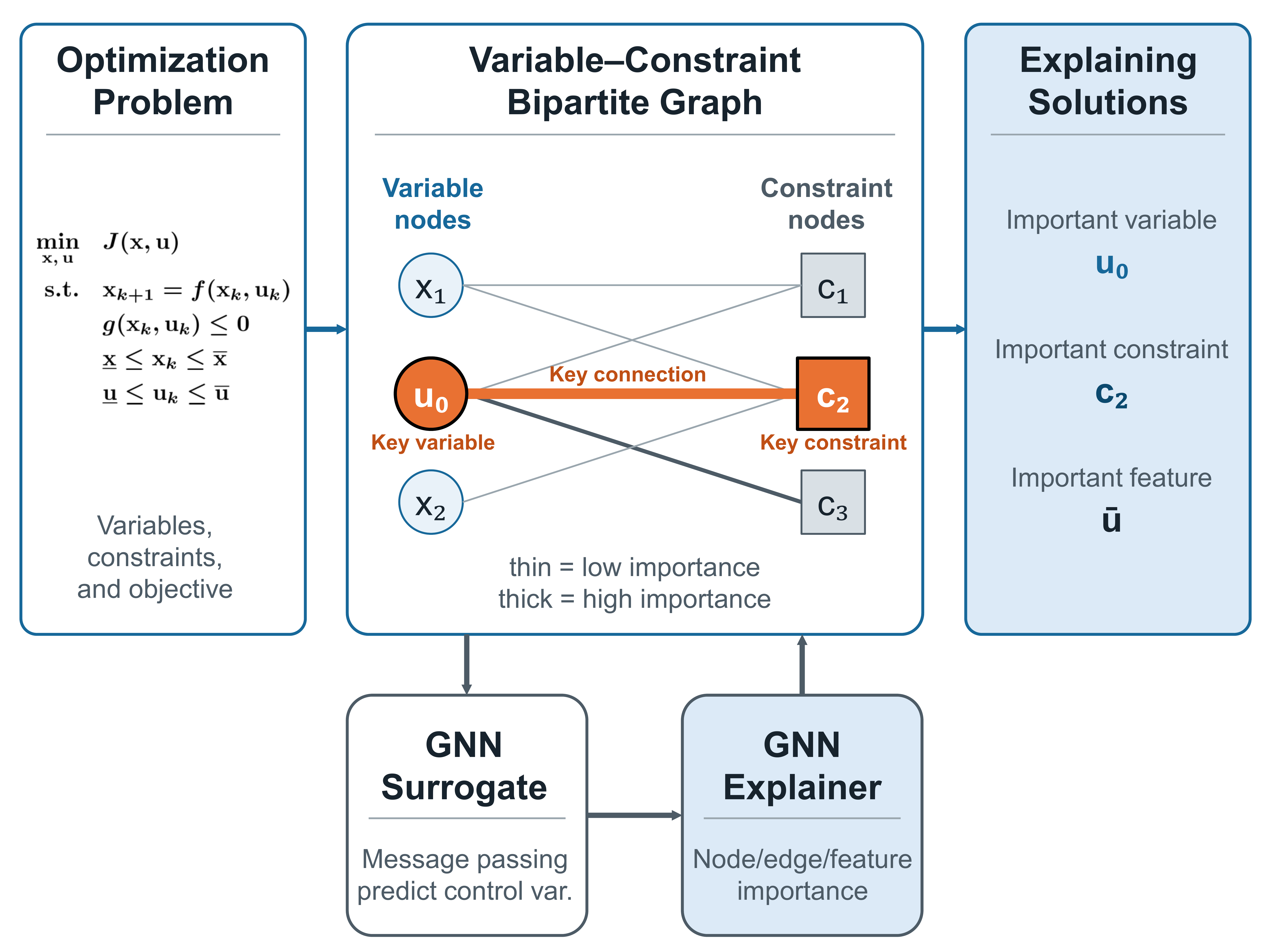}
    \caption{Proposed framework for learning to explain solutions of optimal control problems.}
    \label{fig:framework}
\end{figure}

In this paper, we propose a novel approach to explain the solution of an optimization problem by combining explainable artificial intelligence and graph neural networks (GNN). The proposed approach is presented in Fig.~\ref{fig:framework} and has two components. First, a GNN model is trained to predict the solution of an optimal control problem using a graph representation of the problem as an input. This representation captures the structural coupling between the variables and constraints of the control problem and the functional coupling, i.e., the bounds and domain of the variables and constraint types. Given the trained GNN model, graph explanation algorithms seek a subgraph, i.e., a subset of nodes, edges, and features, which, if passed through the trained GNN, will lead to the same prediction as the original graph. We use the identified subgraph to explain the prediction of the optimal solution using GNNs.

We use the proposed approach to explain the solution of optimal control problems arising in the operation of an isothermal continuous stirred tank reactor. For a given initial condition, set-point profile, and ramping rates, the GNN model predicts the inlet flow rate over the entire time horizon. Application of the proposed framework to an instance of the optimal control problem shows that the most important constraints for predicting the value of the manipulated variables at the optimal solution are ramping constraints, some of which are also active at the optimal solution. The rest of the paper is organized as follows: In Section~\ref{sec: predict u} we present the application of GNNs for predicting the solution of optimal control problems; in Section~\ref{sec: framework} we present the application of explainable AI algorithms for obtaining the most important variables, constraints, and parameters of the optimization problem, and in Section~\ref{sec:case study} we present the application of the proposed framework to a case study.

\section{Learning to predict the solution of an optimal control problem with GNNs} \label{sec: predict u}
\subsection{Dynamic optimization problem}
We consider a dynamical system with ${N_{x}}$ state variables, $x \in \mathbb{R}^{{N_{x}}}$, and ${N_{u}}$ manipulated variables, $u \in \mathbb{R}^{{N_{u}}}$. We define $\mathcal{N}_{x} = \{1,\dots, {N_{x}}\}$ as the set of state variables and $\mathcal{N}_{u} = \{1,\dots, {N_{u}}\}$ as the set of manipulated variables. The dynamic behavior of the system is described by a set of ordinary differential equations
\begin{equation*}
    \dot{x} = f(x(t),u(t)),
\end{equation*}
where $f: \mathbb{R}^{N_{x} + N_{u}} \mapsto \mathbb{R}^{N_{x}}$ is a smooth vector function. Given a desired set-point $x^{sp}(t)$, we seek the temporal profile of the manipulated variables $u(t)$ that minimizes the difference between the state variables and their  over a time horizon $[0,{T}]$. We formulate the optimal control problem as a nonlinear programming problem by discretizing the time domain into ${N_{T}}$ equidistant points. Under this discretization, we define $x_{kt}$ as the value of state variable $k$ at time point $t$ and $u_{kt}$ as the value of manipulated variable $k$ at time point $t$. The discretized optimal control problem is 
\begin{equation}
\label{eq:dynamic_optimization}
\begin{aligned}
\min \quad & \sum_{k\in \mathcal{N}_{x}} \sum_{t \in \mathcal{T}} (x_{kt}-x_{kt}^{sp})^{2} \\
\text{s.t.}\quad & h(\{x_{kt}\},\{u_{kt}\}) =  0 \\
& x_{t} - x_{t-1} \leq  {\Delta x}^{\mathrm{ub}}  \ \forall t  \in \mathcal{T} \setminus \{0\}\\
& x_{t} - x_{t-1} \geq  {\Delta x}^{\mathrm{lb}}  \ \forall t  \in \mathcal{T} \setminus \{0\}\\
& u_{t} - u_{t-1} \leq  {\Delta u}^{\mathrm{ub}}  \ \forall t  \in \mathcal{T} \setminus \{0\}\\
& u_{t} - u_{t-1} \geq  {\Delta u}^{\mathrm{lb}}  \ \forall t  \in \mathcal{T} \setminus \{0\}\\
& {\Delta u}^{\mathrm{lb}} \leq u_{0} - u^{\rm{init}} \leq  {\Delta u}^{\mathrm{ub}} \\
& x_{0} = x^{\rm{init}}\\
& x^{\mathrm{lb}}\leq x_t\leq x^{\mathrm{ub}}, \ \forall t  \in \mathcal{T}\\
& u^{\mathrm{lb}}\leq u_t\leq u^{\mathrm{ub}}, \ \forall t  \in \mathcal{T}, 
\end{aligned}
\end{equation}
where $x^{\rm{init}}$ is the initial state of the system, $u^{\rm{init}}$ is the initial value of the manipulated variables, $x^{\mathrm{lb}}$ and $x^{\mathrm{ub}}$ are the lower and upper bounds for the state variables,  $u^{\mathrm{lb}}$ and $u^{\mathrm{ub}}$ are the bounds for the manipulated variables, ${\Delta x}^{\mathrm{lb}}$, ${\Delta x}^{\mathrm{ub}}$, ${\Delta u}^{\mathrm{lb}}$, ${\Delta u}^{\mathrm{ub}}$ are the ramping rate bounds for the state and manipulated variables, and $\mathcal{T}$ is the set of time points. The equality constraints $h$ capture the discretized differential equations. We denote $N_{c}$ as the total number of constraints. 

\subsection{Graph representation of the optimization problem}
We represent the optimal control problem in Eq.~\ref{eq:dynamic_optimization} as a variable--constraint bipartite graph \cite{allman2019decode, mitrai2022stochastic} with two sets of nodes: $\mathcal{V}_{\rm{var}}$ which represents the variables ($|\mathcal{V}_{\rm{var}}| = N_{T}(N_{x}+N_{u})=N_{v}$) and $\mathcal{V}_{\rm{con}}$ which represents the constraints ($|\mathcal{V}_{\rm{con}}| = N_{c}$). An edge $e_{ij}$ between variable $i \in \mathcal{V}_{\rm{var}}$ and constraint $j \in \mathcal{V}_{\rm{con}}$ denotes the presence of variable $i$ in constraint $j$, i.e., $e_{ij}=1$ if variable $i$ is present in constraint $j$ and zero otherwise. We denote the set of edges as $\mathcal{E} = \{ e_{ij}: i \in \mathcal{V}_{\rm{var}}, j \in \mathcal{V}_{\rm{con}}\}$, the number of edges as $N_{E}$, and define $\mathcal{V} = \mathcal{V}_{\mathrm{var}} \cup \mathcal{V}_{\mathrm{con}}$ as the set of all the nodes in the bipartite graph ($|\mathcal{V}|=N$). The coupling between the variable and constraint nodes is encoded in the adjacency matrix $A \in \mathbb{R}^{N_{v} \times N_{c}}$ with $A_{ij} = e_{ij}$. 

We use $N_{\phi}$ features, for each node, to capture information about variables and constraints. Such nodal and edge features are commonly used to enrich the graph representation of an optimization problem such that downstream ML-related tasks can leverage extra information about the problem \cite{gasse2019exact, mitrai2024takingpart1}. Specifically, for the nodes representing the $x$ variables, the features are
\begin{equation*}
\phi^{x}_i=\left[\mathbf{e}_i,x_{i}^{\mathrm{lb}},x_{i}^{\mathrm{ub}},{\Delta x}^{\mathrm{lb}}, {\Delta x}^{\mathrm{ub}},x^{\rm{sp}}_v,x^{\mathrm{init}},\tau_i\right].
\label{eq:node_features}
\end{equation*}
These features capture the  variable bounds ($x_{i}^{\mathrm{lb}}$ and $x_{i}^{\mathrm{ub}}$), lower and upper ramping rates (${\Delta x}^{\mathrm{lb}}$, ${\Delta x}^{\mathrm{ub}}$),  ($x^{\rm{sp}}$), initial value ($x^{\mathrm{init}}$), and normalized time index $\tau_i = t/(T-1)$. The first feature $\mathbf{e}_{i}\in\{0,1\}^5$ is a one-hot vector that distinguishes node types: state variable, manipulated variable, mass balance constraint, and ramping up/down constraints. To improve learning, we normalize these features as follows
\begin{equation*}
\phi^{x}_i=\left[\mathbf{e}_i,\frac{x_{i}^{\mathrm{lb}}}{x_{i}^{\mathrm{ub}}},1, \frac{{\Delta x}^{\mathrm{lb}}}{{\Delta x}^{\mathrm{ub}}},1, x_{i}^{\rm{sp}},\frac{x^{\mathrm{init}}_{i}}{x_{i}^{\mathrm{ub}}},\tau_i\right].
\label{eq:node_features}
\end{equation*}
Similarly, for the variable nodes representing the manipulated variables, the features are
\begin{equation}
\phi_i^{u} =\left[\mathbf{e}_i, \frac{u^{\mathrm{lb}}}{u^{\mathrm{ub}}}, 1, \frac{{\Delta u}^{\mathrm{lb}}}{{\Delta u}^{\mathrm{ub}}},1, 0,\frac{u^{\mathrm{init}}_{i}}{u^{\mathrm{ub}}},\tau_i\right],
\label{eq:node_features}
\end{equation}
Note that the feature for the manipulated variable set point is set to zero. The $x^{\mathrm{init}}$ and $u^{\mathrm{init}}$ features for the nodes representing $x_{t}$ and $u_{t}$ are set equal to zero for $t>0$. Finally, for the nodes representing constraints, the features are
\begin{equation}
\phi_i^{c} =\left[\mathbf{e}_i,0,0,0,0,0,0,\tau_i\right].
\label{eq:node_features}
\end{equation}
Overall, the nodal features are captured in the feature matrix $\Phi$ which is equal to 
\begin{equation}
\Phi = \left[\Phi_x^\top \;\, \Phi_u^\top \;\, \Phi_c^\top\right]^\top \in \mathbb{R}^{|\mathcal{V}| \times N_\phi},
\label{eq:feature_matrix}
\end{equation}
where the $i$-th rows of $\Phi_x$, $\Phi_u$, and $\Phi_c$ are $\phi_i^{x}$, $\phi_i^{u}$, and $\phi_i^{c}$, respectively. Under this representation, an instance of the optimization problem in Eq.~\ref{eq:dynamic_optimization} is represented by a graph $\mathcal{G}(A,\Phi)$ where the adjacency matrix $A$ captures the structural coupling among the variables and the constraints, and $\Phi$ captures information such as the bounds, ramping rates, s, and initial values.

\subsection{Predicting the optimal solution using GNNs} \label{sec: GNN}
The variable--constraint bipartite graph $\mathcal{G}(A,\Phi)$ is used as an input to a graph neural network that predicts the optimal values of the manipulated variables. We formulate the GNN training as a node regression problem: the input is the graph representation with the associated features, and the output for each node $k$ representing a manipulated variable is the predicted optimal solution $\hat{u}_{kt}$. The prediction of the optimal solution is performed in three steps. 

First, the node features are updated based on neighbors' features using message passing. Specifically, a graph convolutional neural network (GCN) \cite{kipf2016semi} with $N_{l}$ layers and $N_{h}$ hidden dimension are used to update the features as follows
\begin{equation}
\begin{aligned}
h_i^{(\ell+1)}&=\sigma\left(\sum_{j \in \mathcal{N}(i) \cup\{i\}}\frac{1}{\sqrt{\tilde d_i\tilde d_j}}\mathbf{W}^{(\ell)}h_j^{(\ell)} +\mathbf{b}^{(\ell)}\right),
\end{aligned}
\end{equation}
where $h_i^{(0)} = \phi_i$ (the features of node $i$), 
$h_{i}^{(\ell)} \in \mathbb{R}^{N_h}$ is the embedding of node $i$ at layer $\ell$ ($N_{h}$ is the hidden dimension), $\mathcal{N}(i)$ denotes the neighbors of node $i$, $\tilde{d}_i$ and $\tilde{d}_j$ are the degrees of node $i$ and $j$, $\mathbf{W}^{(\ell)}$ is a learning weight matrix, $\textbf{b}^{(\ell)}$ is the bias, and $\sigma(\cdot)$ is the activation function. After $N_{l}$ message passing layers, the features of node $i$ are denoted as $h_{i}^{N_{l}}$.

In the second step, we predict the optimal value for each manipulated variable by considering both the learned representation for the variable, $h_{i}^{N_{l}}$, and a graph-level feature $H^{G} \in \mathbb{R}^{N_{h}}$ obtained using a global mean pooling layer as presented below
\begin{equation}
H^{G}_{k}=\frac{1}{|\mathcal{V}|} \sum_{i \in \mathcal{V}} h_{i,k}^{N_{l}} \ \ \forall k=1,\dots, N_{h}.
\end{equation}
This graph-level feature captures information about the entire optimization problem and augments each node's representation before prediction. Specifically, the local node embedding and global graph embedding are concatenated 
\begin{equation}
z_{i} = [h_i^{N_{l}} , \| , H^{G}],
\end{equation}
with $\|$ denoting vector concatenation, and the resulting vector is used as an input to a Multi-layer Perceptron (MLP) to generate a scalar output $\hat{u}_{i}$ as shown below
\begin{equation}
    \hat{u}_{i} = \mathrm{MLP}( z_{i}; w_{\mathrm{MLP}}),
\end{equation}
where $w_{\mathrm{MLP}}$ are the weights of the MLP layer.

\subsection{Training the GNN model}
We generate $N_{\mathrm{data}}$ optimization instances by varying the set-point, initial state and manipulated variables, and ramping rates. For each optimization problem $p$, we obtain the graph representation $\mathcal{G}(A_{p}, \Phi_{p})$ and solve it using a nonlinear optimization solver. We denote $u^{*}_{pkt}$ as the optimal value of manipulated variable $k$ at time $t$ for training data point $p$, i.e., optimization problem instance, and $\textbf{u}_{p}^{*}$ as the vector of the optimal manipulated variables for data point $p$. Overall, the training data set is $\mathcal{D}_{\mathrm{data}} = \{(A_{p},\Phi_{p}), \textbf{u}^{*}_{p}\}_{p=1}^{N_{data}}$ and the learning task is
\begin{equation}
    \min_{\textbf{W}^{(\ell)}, \textbf{b}^{(\ell)}, w_{\mathrm{MLP}}} \frac{1}{N_{\mathrm{data}}} \sum_{p=1}^{N_{\mathrm{data}}} \sum_{k \in \mathcal{N}_{u}} \sum_{t \in \mathcal{T}} \left( \frac{\hat{u}_{pkt} - u_{pkt}^{*}}{u^{\mathrm{ub}}} \right)^2.
\end{equation}

\section{Learning to Explain the Predicted Solution} \label{sec: framework}

Given an optimization problem, its graph representation $\mathcal{G}(A,\Phi)$, and a trained GNN model, we are interested in identifying a subset of the variables, constraints, variable--constraint connections, and parameters that affect the prediction of the GNN model the most when predicting the optimal value of the manipulated variable $\bar{u}_{kt} = \mathrm{GNN}(A,\Phi)$. Mathematically, this is equivalent to finding a subset of the nodes, edges, and features, i.e., a subgraph $\mathcal{G}_{s}(A_{s},\Phi_{s})$ which, if used as input to the GNN model, will lead to the same (or similar) prediction. We will refer to this subgraph as the explanation graph. We use the GNNExplainer algorithm from \cite{ying2019gnnexplainer} to explain the prediction of the manipulated variable. 

We define a binary variable $m_{ij}$ which is equal to one if the edge between nodes $i$ and $j$ is part of the explanation graph and zero otherwise. We also define a binary variable $\hat{m}_{ik}$ which is equal to one if feature $k$ at node $i$ is part of the explanation. Given these variables, the adjacency and feature matrices of the explanation graph are $\hat{A} = A \odot M$ and $\hat{\Phi} = \Phi \odot \hat{M}$, respectively, and $M \in \{0,1\}^{N_{v} \times N_{c}}$ and $\hat{M} \in \{0,1\}^{N \times N_{\phi}}$ are matrices obtained by stacking the $m$ and $\hat{m}$ variables. The explanation task is a combinatorial optimization problem presented below
\begin{equation*} 
\begin{aligned}
    \min_{M , \hat{M}} \ (u_{kt}^{\mathrm{pred}} - \bar{u}_{kt})^{2} + \lambda_{E} \sum_{i=1}^{N_{v}} \sum_{j=1}^{N_{c}} M_{ij} + \frac{\lambda_{F}}{N_{\phi}N} \sum_{i=1}^{N} \sum_{k=1}^{N_{\phi}} \hat{M}_{ik},  
\end{aligned}
\end{equation*}
where $u^{\mathrm{pred}} = \mathrm{GNN}(\hat{A},\hat{\Phi})$, and $\lambda_{E}$, $\lambda_{F}$ are weights. To improve tractability, the binary variables are relaxed and entropy penalties are added, resulting in the following continuous explanation task \cite{ying2019gnnexplainer}
\begin{equation} \label{eq: explainer}
\begin{aligned}
    \min_{M , \hat{M}} \ & (u_{kt}^{\mathrm{pred}} - \bar{u}_{kt})^{2} \\
    & + \lambda_{E} \sum_{i=1}^{N_{v}} \sum_{j=1}^{N_{c}} M_{ij} + \frac{\lambda_{F}}{N_{\phi}N} \sum_{i=1}^{N} \sum_{k=1}^{N_{\phi}} \hat{M}_{ik} \\
    & + \frac{\beta_{E}}{|\mathcal{E}|} \sum_{i=1}^{N_{v}} \sum_{j=1}^{N_{c}} h(M_{ij}) + \frac{\beta_{F}}{N_{\phi} N} \sum_{i=1}^{N}\sum_{k=1}^{N_{\phi}} h(\hat{M}_{ik}),
\end{aligned}
\end{equation}
where $h(\cdot)$ is the entropy function, and $\beta_{E}$, $\beta_{F}$ are weights. The first term is the difference between the GNN prediction using the explanation and the original graphs. The second and third terms encourage small explanations, i.e., a few nodes and edges to exist in the explanation graph, and the last two terms encourage the binary variables to be either zero or one. This problem is solved to local optimality using gradient-based methods \cite{ying2019gnnexplainer}.

\begin{figure}[!htbp]
    \centering
    \includegraphics[width=0.9\columnwidth]
    {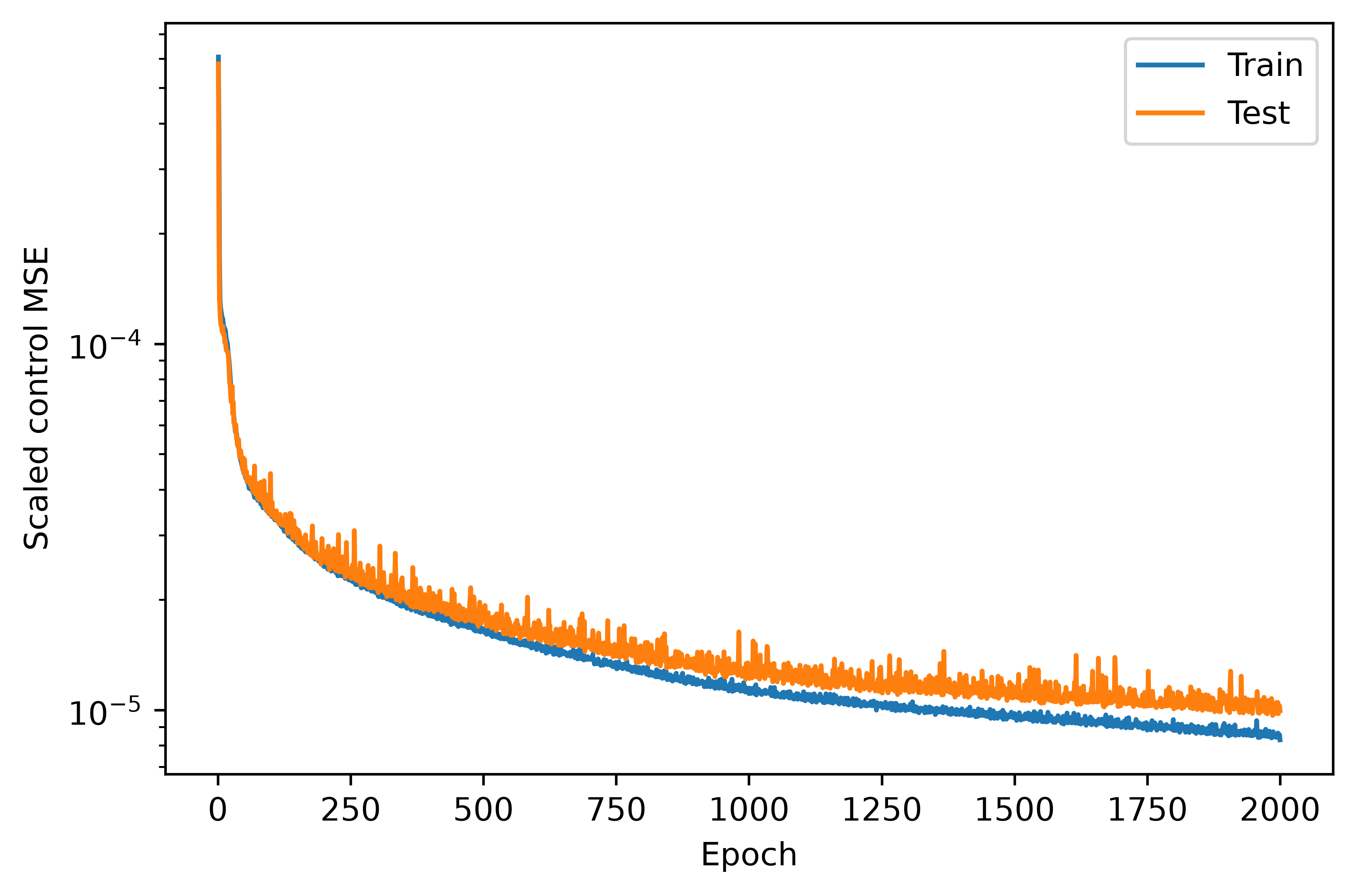}
    \caption{Training and test loss of the GNN surrogate.}
    \label{fig:loss}
\end{figure}

\begin{figure}[!htbp]
    \centering
    \begin{subfigure}[t]{0.9\columnwidth}
        \centering
        \includegraphics[width=\linewidth]
        {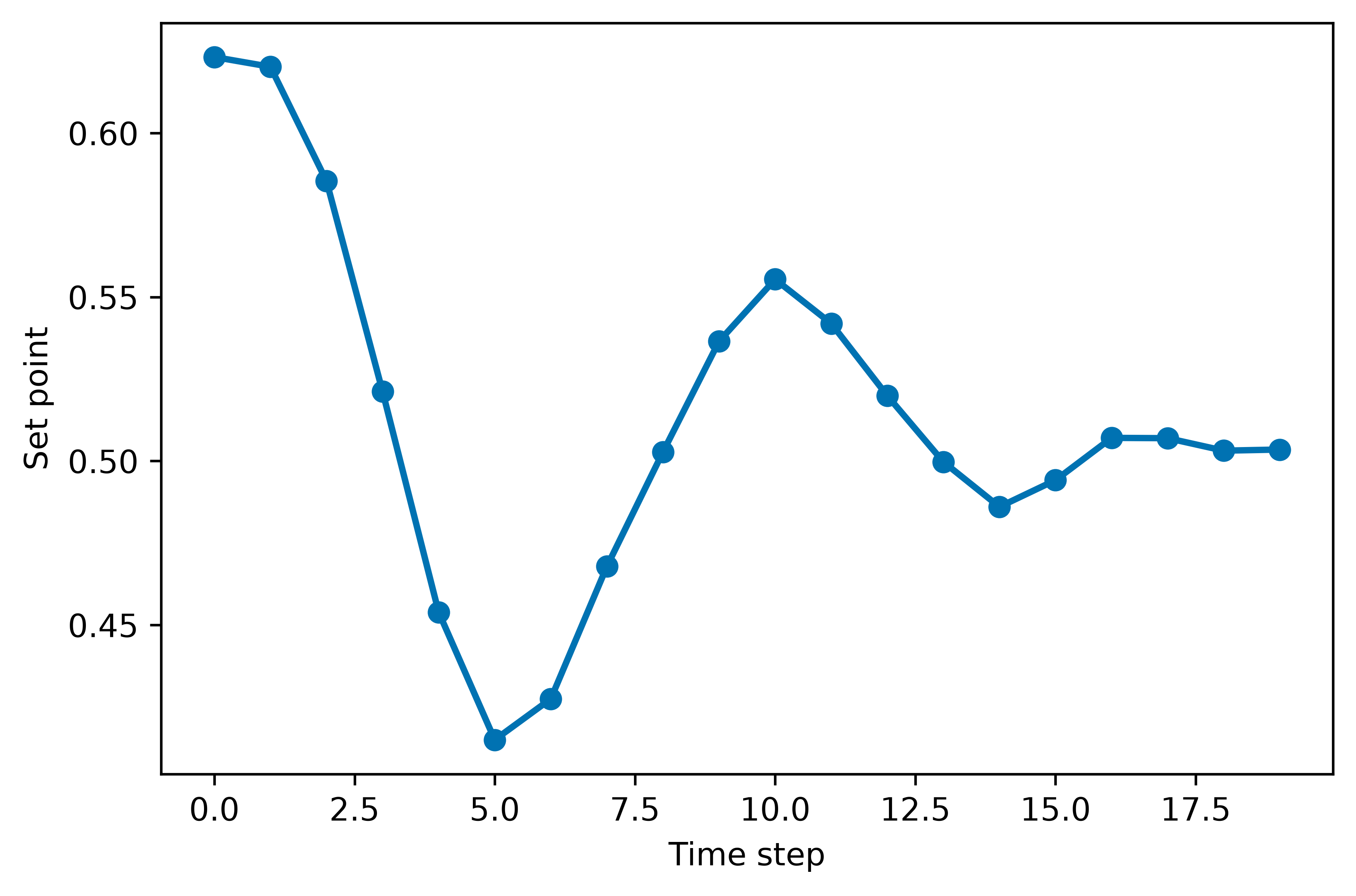}
    \end{subfigure}
    \begin{subfigure}[t]{0.9\columnwidth}
        \centering
        \includegraphics[width=\linewidth]
        {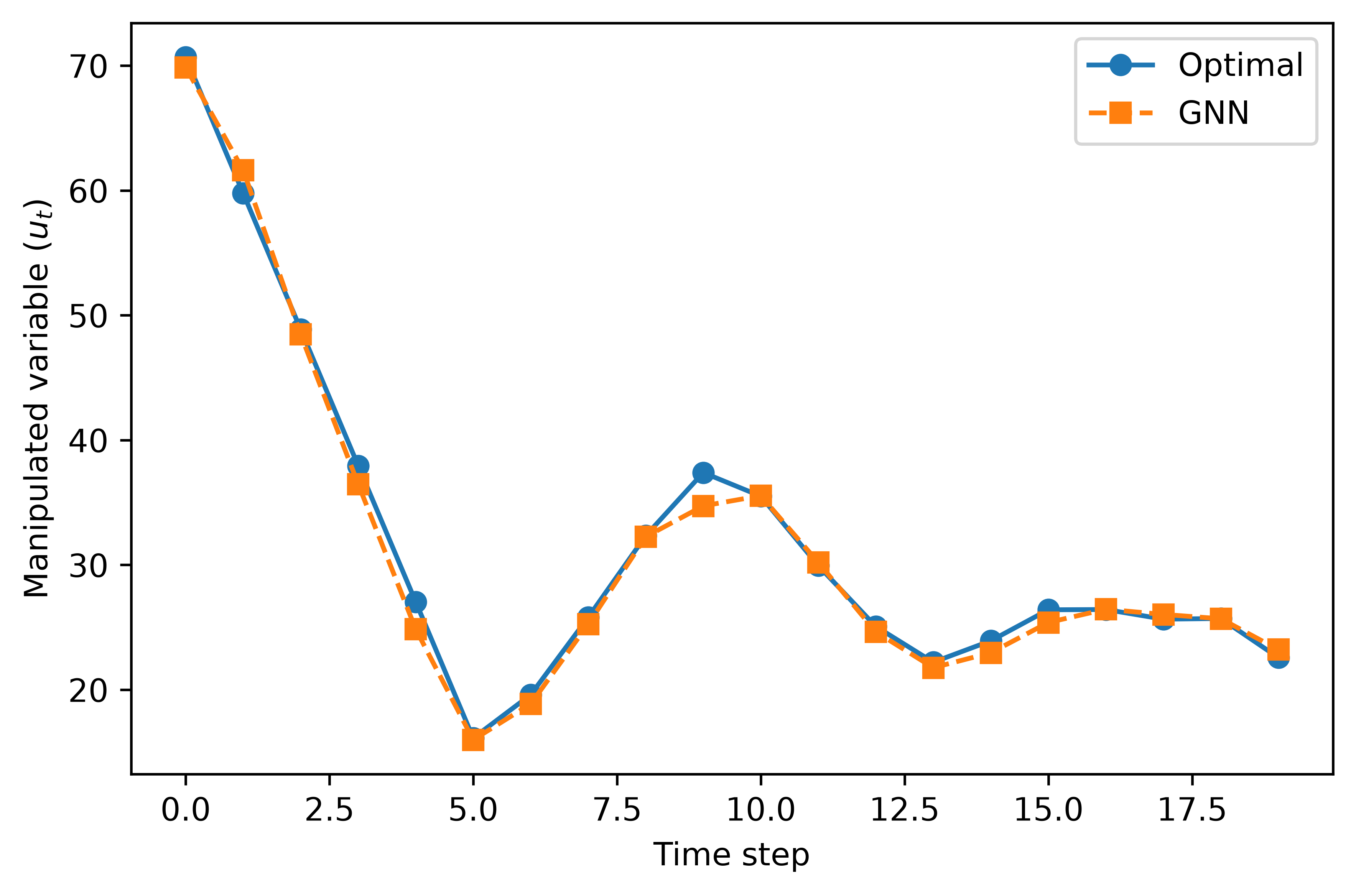}
    \end{subfigure}
    \caption{Set-point profile (top) and GNN prediction (bottom) for the test graph.}
    \label{fig:testgraph}
\end{figure}

\section{Computational Results} \label{sec:case study}
\subsection{System description and data generation}
We consider the optimal control of an isothermal continuous stirred-tank reactor (CSTR) where an irreversible reaction $A \rightarrow 3B$ occurs. We define $x(t)$ as the reactant concentration and $u(t)$ as the inlet flow rate. The dynamic behavior of the system is described by the following equation
\begin{equation}
    \frac{dx(t)}{dt} = \frac{u(t)}{V}(x_{\mathrm{feed}} - x(t)) - k x(t)^{3},
\end{equation}
where $V=50 L$ is the  reactor volume, $x_{\mathrm{feed}}=1 mol/L$ is the inlet concentration, and $k=2 L^2/(min \ mol^2)$ is the reaction constant. For a given set-point profile for the concentration $x^{sp}$, ramping rates $(\Delta u^{\mathrm{lb}}, \Delta u^{\mathrm{ub}})$, and initial conditions $(x^{\mathrm{init}}, u^{\mathrm{init}})$, the optimal profile of the manipulated variables is found by solving the following problem
\begin{equation}
\label{eq:CSTR_dynamic_optimization}
\begin{aligned}
\min_{x_{t},u_{t}} \ \
& \sum_{t\in\mathcal{T}} (x_{t}-x_{t}^{\rm sp})^2 \\
\text{s.t.}\ \ & x_{t+1} = x_{t} + \Delta t \big( \frac{u_{t}(1-x_{t})}{V} - k x_{t}^{3}\big) \ \  \forall t \in \hat{\mathcal{T}} \\
& (u_{t}-u_{t-1})/\Delta t \leq  \Delta u^{\mathrm{ub}} \ \  \forall t \in \mathcal{T}\setminus\{0\} \\
& (u_{t}-u_{t-1})/\Delta t \geq  \Delta u^{\mathrm{lb}} \ \ \forall t \in \mathcal{T}\setminus\{0\}\\
& x_{0} = x^{\rm{init}}\\
&u^{\mathrm{init}} +\Delta u^{\mathrm{lb}}\Delta t \leq u_{0} \\
&u_{0} \leq u^{\mathrm{init}} +\Delta u^{\mathrm{ub}}\Delta t \\
& x^{\mathrm{lb}} \leq x_t \leq x^\mathrm{ub} \ \ \forall t \in \mathcal{T} \\
& u^{\mathrm{lb}} \leq u_t \leq u^\mathrm{ub} \ \ \forall t \in \mathcal{T},
\end{aligned}
\end{equation}
where $\hat{\mathcal{T}} = T \setminus \{N_{T}\}$, $\Delta t=0.5 \min$, ${N_{T}}=20$, $x^{\mathrm{lb}}=0 \ mol/L$, $x^{\mathrm{ub}}=3 \ mol/L$, $u^{\mathrm{lb}}=0 \ L/min$, $u^{\mathrm{ub}}=800 \ L/min$. 
\begin{figure*}[h]
    \centering
    \includegraphics[width=\textwidth]
    {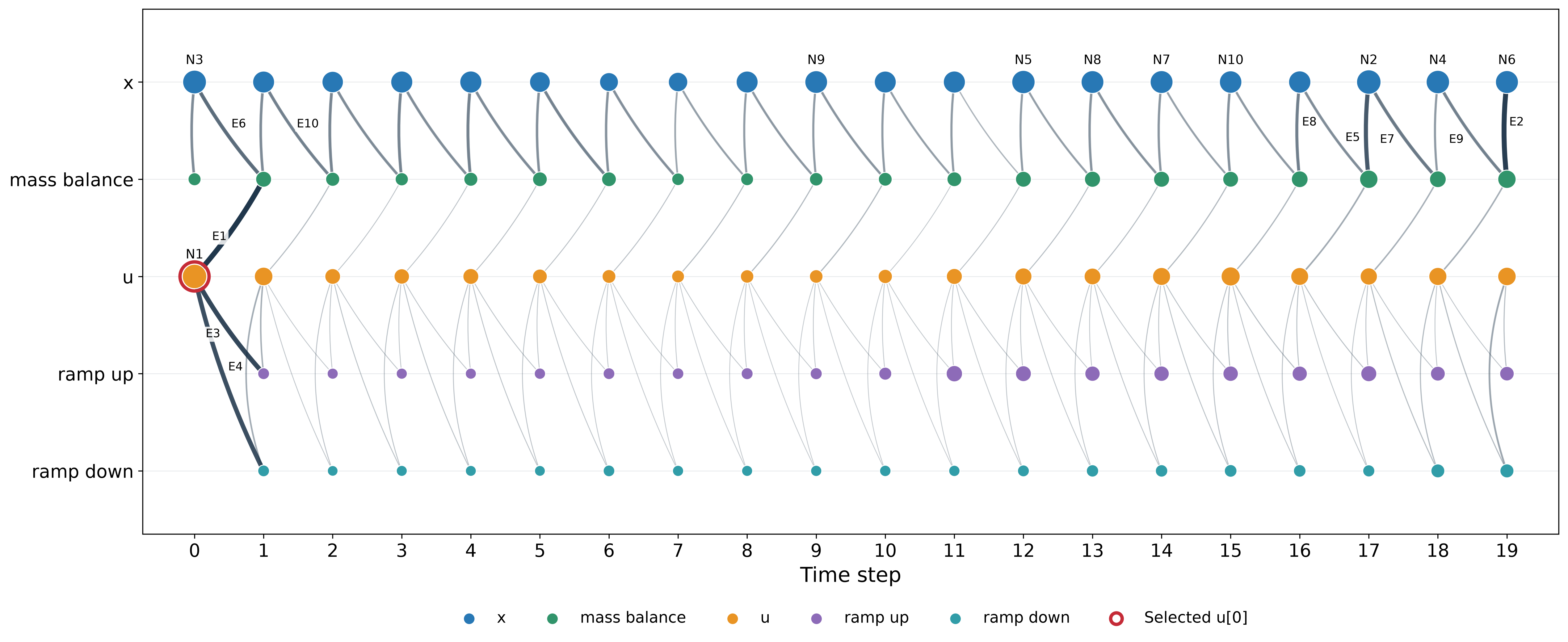}
    \caption{Node and edge importance for the GNN prediction of the first manipulated variable $u_0$ in the test graph. The thickness of an edge and area of a node are proportional to their importance. The label $N{(\cdot)}$ denotes the rank of the importance.}
    \label{fig:explaining_u0}
\end{figure*}
We generate $10^{4}$ optimization problem instances by sampling $x^{\rm{init}}\sim U(0.2,0.8)$, $ u^{\rm{init}}\sim U(10,100)$, $-\Delta u^\mathrm{lb} = \Delta u^\mathrm{ub}  \sim U(10,40)$, and $x^{\rm{sp}}=0.5+0.1 \mathrm{GRF}(l=0.5)$ with $\mathrm{GRF}$ denoting a Gaussian random field with length scale equal to 0.5. The optimization problems are implemented in Pyomo \cite{hart2017pyomo} and solved through GAMS with CONOPT \cite{gams_software}. We use 85\% of the problems for training and the rest for testing. 

\subsection{GNN training and prediction performance}
The architecture is implemented in PyTorch Geometric \cite{fey2019fastgraphrepresentationlearning} with ReLU activation, and we use GCN \cite{kipf2016semi} layers with 64 hidden neurons. We train the GNN using Adam \cite{kingma2014adam} with a learning rate equal to $10^{-2}$ for $2 \times 10^{3}$ iterations and a batch size equal to $50$. The training and testing errors are presented in Fig.~\ref{fig:loss}. The testing normalized mean squared error for the trained GNN model is $9.96\times10^{-6}$ and Fig.~\ref{fig:testgraph} presents the GNN prediction for an optimization problem from the test case. For this specific case, the mean absolute error between the optimal profile of the manipulated variables and the one predicted by the GNN model throughout the time horizon is $0.75$ and the mean relative error is $2.4\%$.

\subsection{Explaining optimal solutions}
Given the trained GNN model, we use the proposed approach to explain the predicted optimal solution for the optimization problem instance presented in Fig.~\ref{fig:testgraph}. For GNNExplainer, we use Adam for $10^{3}$ iterations with a learning rate of $0.03$. The weights are set equal to $\lambda_{E}=\lambda_{F} = 3\times10^{-4}$ and $\beta_{E} = \beta_{F} = 3\times10^{-5}$. The average time required to obtain an explanation for a single control action was approximately 2 seconds.

We first evaluate the difference in the predicted values of the manipulated variables when the explanation and original graphs are used as input to the GNN. We define the absolute fidelity error as $\epsilon_t^{\mathrm{fid}} =|u_{t}^{\mathrm{pred}}-\hat{u}_t|$, where $\hat{u}_{t}$ is the original GNN prediction and $u_{t}^{\mathrm{pred}}$ is the prediction obtained using the explanation graph. For the test example, the mean absolute fidelity error is 0.3392 and the maximum absolute fidelity error is 0.8566 at $t=4$. These low fidelity errors show that the explanation graph produces predictions similar to the original graph.
\begin{table}[t]
\centering
\caption{Top-5 variable-constraint edges for the test graph.}
\label{tab:top_edges}
\small
\begin{tabular}{l c l}
\hline
\textbf{Edge} &
\textbf{Score} &
\textbf{Status} \\
\hline
$u_{0}$ -- Mass balance $(t=1)$   & 1.0000 & Equality \\
$x_{19}$ -- Mass balance $(t=19)$ & 0.9568 & Equality \\
$u_{0}$ -- Ramping up $(t=0)$    & 0.8957 & Inequality-Inactive \\
$u_{0}$ --  Ramping down $(t=0)$ & 0.8483 & Inequality-Active \\
$x_{17}$ -- Mass balance $(t=17)$ & 0.7853 & Equality \\
\hline
\end{tabular}
\end{table}

Fig.~\ref{fig:explaining_u0} shows the relative importance of the nodes and edges when predicting $u_{0}$. The most important node is $u_{0}$, i.e., the features of the node itself, followed by $x_{17}$, $x_{0}$, $x_{18}$, and $x_{12}$. The top-5 edges are tabulated in Table~\ref{tab:top_edges}. The score refers to the values of the binary variables $M_{ij}$ obtained after solving the continuous relaxation of the explainability task (Eq.~\ref{eq: explainer}). The results show that the edge connecting $u_{0}$ and the mass balance at $t=1$ is the most important; this constraint couples $x_{1}$ and $u_{0}$. In addition, the edges connecting $u_{0}$ to the ramping-up and ramping-down constraints at $t=0$ are also important. We also note that several highly ranked nodes and edges occur near the end of the time horizon. This result could be driven by the GNN architecture where each control prediction uses both its local GCN embedding and the global mean of all node embeddings. Also, it can be caused by solving the continuous relaxation of the explanation task locally. 

\begin{figure}[h]
    \centering
    \includegraphics[width=0.9\columnwidth]
    {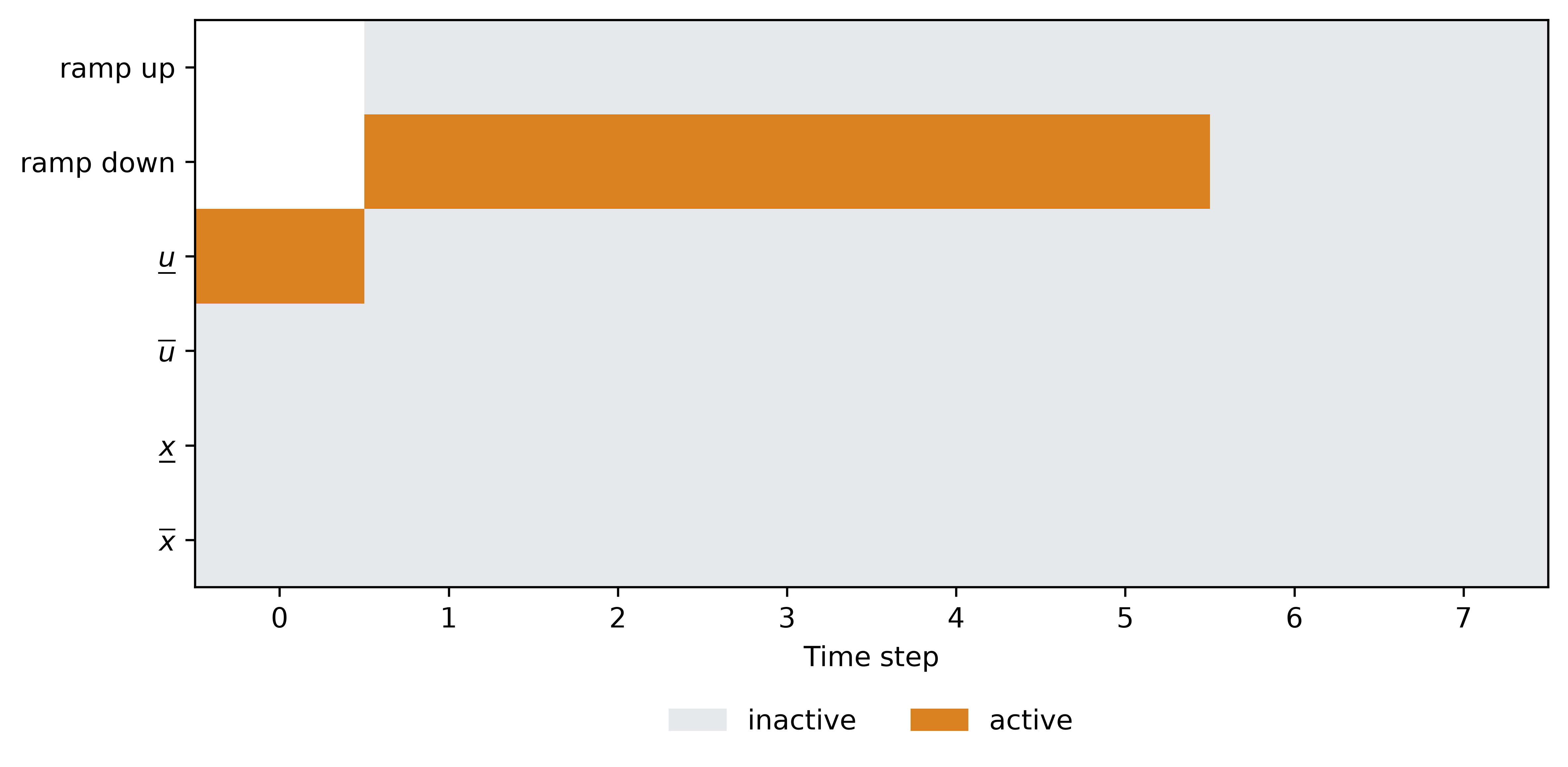}
    \caption{Active/inactive inequalities and variable bounds at the optimal solution for the first eight time steps of the test graph. After time step seven, all inequalities are inactive and not shown.}
    \label{fig:activeConstraint}
\end{figure}

We use the solver output for the specific instance to assess the validity of the explanation. Fig.~\ref{fig:activeConstraint} shows that for the given problem, the ramp-down inequality constraints are active from time step 1 to 5, the lower bound of $u_{0}$ is active, and all remaining inequality constraints and variable bounds are inactive. The explanation from the proposed approach shows partial agreement with the active set of constraints, since the ramping-down constraint for $u_{0}$, which the proposed approach identifies as important, is active at the optimal solution. We note however, that the proposed approach does not explain the solver output, rather it explains the prediction of the GNN model which is trained using the optimal solution obtained from the solver.

\begin{figure}[h]
    \centering
    \includegraphics[width=\columnwidth]
    {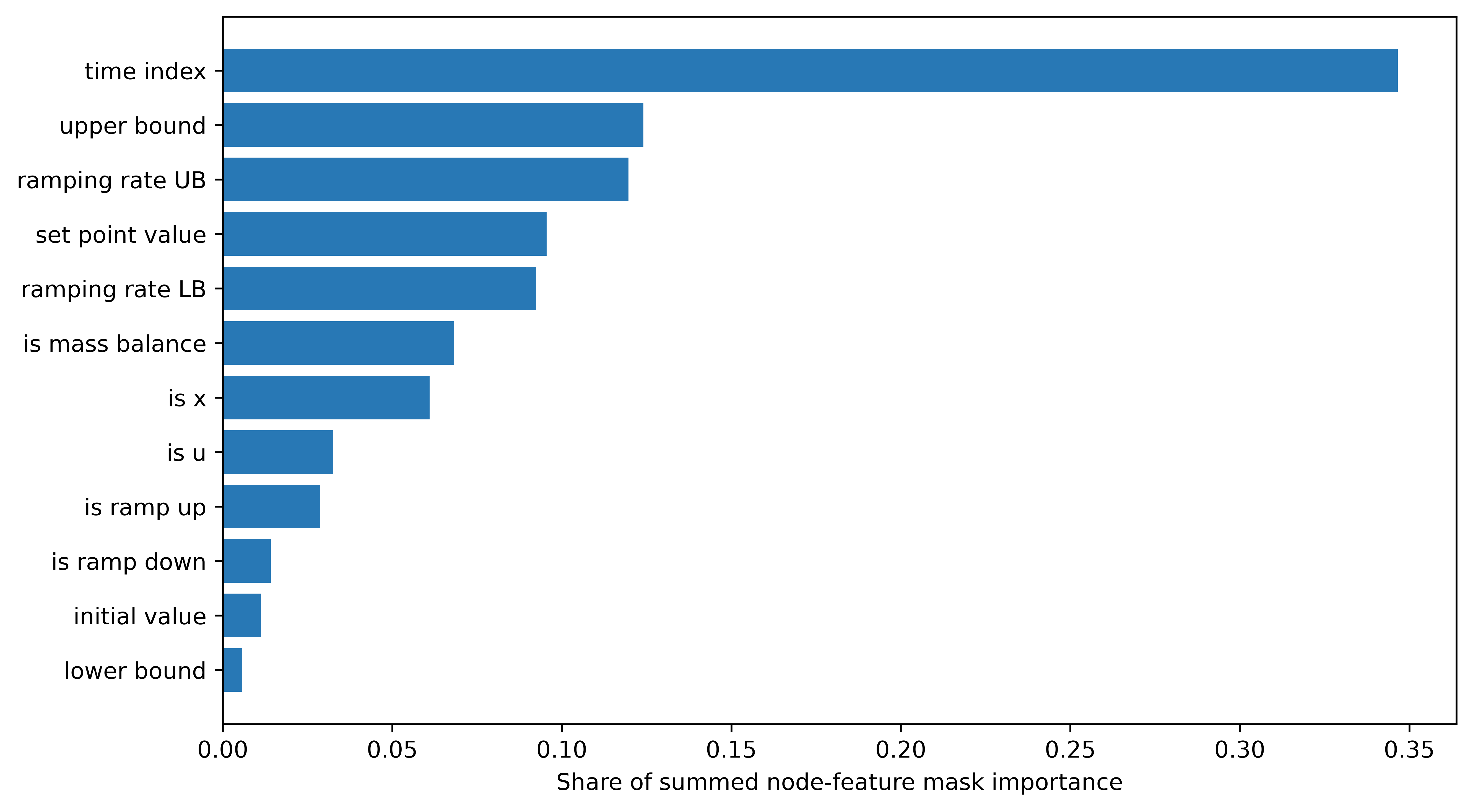}
    \caption{Feature importance across all nodes in the graph to predict the first control action ($u_{0}$) of the test graph.}
    \label{fig:featureImportance}
\end{figure}

Finally, we analyze the importance of each feature when predicting $u_{0}$. The average feature importance across all the nodes is presented in Fig.~\ref{fig:featureImportance}. The results show that the time index is the most important feature, followed by the upper bound and the ramping rate.

\section{Conclusions}

In this paper, we proposed a graph-based explainable AI framework to interpret solutions of optimal control problems predicted with GNNs. The optimal control problem is represented as a variable-constraint bipartite graph, which is the input to a GNN model which predicts the values of the manipulated variables at the optimal solution. The explanation algorithm identifies the most important variables, constraints, and parameters when predicting the optimal solution for each manipulated variable. Application to an isothermal CSTR case study shows that GNNs can accurately predict the optimal values of the manipulated variables. Moreover, the explanation obtained captures the ``local" structure of the optimization problem and some of the identified ``important" inequalities correspond to active constraints in the optimal solution.

Although these results show the promise of the proposed integrated predict-then-explain approach, several limitations exist. First, the proposed approach explains the GNN's predictions, not necessarily the behavior of a numerical solver. Moreover, the explanation task is solved to local optimality, which can affect the quality of the explanation obtained. Although mixed-integer optimization solvers can be used to solve the explanation problem \cite{gaines2025explaining}, their scalability can be limited for large graphs. Finally, we note that the evaluation of an explanation is dependent on human intuition since the ``correct" explanation for an optimization problem is not available.

\addtolength{\textheight}{-12cm}   

\vspace{10pt}
\addcontentsline{toc}{section}{References}
\bibliographystyle{ieeetr}
\bibliography{bib_sample}

@article{fey2019fastgraphrepresentationlearning,
  title   = {Fast Graph Representation Learning with {PyTorch Geometric}},
  author  = {Fey, Matthias and Lenssen, Jan E.},
  journal = {arXiv preprint arXiv:1903.02428},
  year    = {2019}
}

@inproceedings{mitrai2026discovering,
  title={Discovering interpretable piecewise nonlinear model predictive control laws via symbolic decision trees},
  author={Mitrai, Ilias},
  booktitle={2026 American Control Conference (ACC)},
  pages={4248-4253},
  year={2026},
  organization={IEEE}
}

@article{lundberg2017unified,
  title={A unified approach to interpreting model predictions},
  author={Lundberg, Scott M and Lee, Su-In},
  journal={Advances in neural information processing systems},
  volume={30},
  year={2017}
}

@article{gaines2025explaining,
  title={Explaining Graph Neural Networks with mixed-integer programming},
  author={Gaines, Blake B and Zhu, Chunjiang and Bi, Jinbo},
  journal={Neurocomputing},
  volume={639},
  pages={130214},
  year={2025},
  publisher={Elsevier}
}

@inproceedings{ribeiro2016should,
  title={" Why should i trust you?" Explaining the predictions of any classifier},
  author={Ribeiro, Marco Tulio and Singh, Sameer and Guestrin, Carlos},
  booktitle={Proceedings of the 22nd ACM SIGKDD international conference on knowledge discovery and data mining},
  pages={1135--1144},
  year={2016}
}

@INPROCEEDINGS{11615398,
  author={Nassaji, Amin and Mitrai, Ilias and Daoutidis, Prodromos},
  booktitle={2026 American Control Conference (ACC)}, 
  title={Approximate Dynamic Optimization via Deep Neural Operators}, 
  year={2026},
  pages={4254-4259},
  organization={IEEE}
}

@manual{gams_software,
  author       = {{GAMS Development Corporation}},
  title        = {General Algebraic Modeling System (GAMS) Release 54.3.1},
  year         = {2026},
  address      = {Fairfax, VA, USA},
}

@article{ying2019gnnexplainer,
  title={Gnnexplainer: Generating explanations for graph neural networks},
  author={Ying, Zhitao and Bourgeois, Dylan and You, Jiaxuan and Zitnik, Marinka and Leskovec, Jure},
  journal={Advances in neural information processing systems},
  volume={32},
  year={2019}
}

@article{allman2019decode,
  title={De{COD}e: a community-based algorithm for generating high-quality decompositions of optimization problems},
  author={Allman, Andrew and Tang, Wentao and Daoutidis, Prodromos},
  journal={Optim. Eng.},
  volume={20},
  number={4},
  pages={1067--1084},
  year={2019},
  publisher={Springer}
}

@article{hertneck2018learning,
  title={Learning an approximate model predictive controller with guarantees},
  author={Hertneck, Michael and K{\"o}hler, Johannes and Trimpe, Sebastian and Allg{\"o}wer, Frank},
  journal={IEEE Control Syst. Lett.},
  volume={2},
  number={3},
  pages={543--548},
  year={2018},
  publisher={IEEE}
}

@article{mitrai2022stochastic,
  title={Stochastic blockmodeling for learning the structure of optimization problems},
  author={Mitrai, Ilias and Tang, Wentao and Daoutidis, Prodromos},
  journal={AIChE J.},
  volume={68},
  number={6},
  pages={e17415},
  year={2022},
  publisher={Wiley Online Library}
}

@book{hart2017pyomo,
	title={Pyomo -- optimization modeling in {Python}},
	author={Hart, William E and Laird, Carl D and Watson, Jean-Paul and Woodruff, David L and Hackebeil, Gabriel A and Nicholson, Bethany L and Siirola, John D},
	edition={2nd},
	year={2017},
	publisher={Springer}
}

@article{gasse2019exact,
  title={Exact combinatorial optimization with graph convolutional neural networks},
  author={Gasse, Maxime and Ch{\'e}telat, Didier and Ferroni, Nicola and Charlin, Laurent and Lodi, Andrea},
  journal={Advances in Neural Information Processing Systems},
  volume={32},
  year={2019}
}

@article{kingma2014adam,
  title={Adam: A method for stochastic optimization},
  author={Kingma, Diederik P and Ba, Jimmy},
  journal={arXiv preprint arXiv:1412.6980},
  year={2014}
}

@inproceedings{karg2018deep,
  title={Deep learning-based embedded mixed-integer model predictive control},
  author={Karg, Benjamin and Lucia, Sergio},
  booktitle={2018 European Control Conference (ECC)},
  pages={2075--2080},
  year={2018},
  organization={IEEE}
}

@article{kumar2021industrial,
  title={Industrial, large-scale model predictive control with structured neural networks},
  author={Kumar, Pratyush and Rawlings, James B and Wright, Stephen J},
  journal={Comput. Chem. Eng.},
  volume={150},
  pages={107291},
  year={2021},
  publisher={Elsevier}
}

@inproceedings{cauligi2022prism,
  title={{PRISM}: Recurrent Neural Networks and Presolve Methods for Fast Mixed-integer Optimal Control},
  author={Cauligi, Abhishek and Chakrabarty, Ankush and Di Cairano, Stefano and Quirynen, Rien},
  booktitle={Learning for Dynamics and Control Conference},
  pages={34--46},
  year={2022},
  organization={PMLR}
}

@article{kipf2016semi,
	title={Semi-supervised classification with graph convolutional networks},
	author={Kipf, Thomas N and Welling, Max},
	journal={arXiv preprint arXiv:1609.02907},
	year={2016}
}

@article{chen2024diagnosing,
  title={Diagnosing infeasible optimization problems using large language models},
  author={Chen, Hao and Constante-Flores, Gonzalo E and Li, Can},
  journal={INFOR: Information Systems and Operational Research},
  volume={62},
  number={4},
  pages={573--587},
  year={2024},
  publisher={Taylor \& Francis}
}

@inproceedings{aigner2024framework,
  title={A framework for data-driven explainability in mathematical optimization},
  author={Aigner, Kevin-Martin and Goerigk, Marc and Hartisch, Michael and Liers, Frauke and Miehlich, Arthur},
  booktitle={Proc. AAAI Conf. Artif. Intell.},
  volume={38},
  number={19},
  pages={20912--20920},
  year={2024}
}

@article{rathi2024enhancing,
  title={Enhancing explainability of stochastic programming solutions via scenario and recourse reduction},
  author={Rathi, Tushar and Gupta, Rishabh and Pinto, Jose M and Zhang, Qi},
  journal={Optim. Eng.},
  volume={25},
  number={2},
  pages={795--820},
  year={2024},
  publisher={Springer}
}

@article{greenberg1993analyze,
  title={How to analyze the results of linear programs—Part 2: Price interpretation},
  author={Greenberg, Harvey J},
  journal={Interfaces},
  volume={23},
  number={5},
  pages={97--114},
  year={1993},
  publisher={INFORMS}
}

@inproceedings{korikov2023objective,
  title={Objective-Based Counterfactual Explanations for Linear Discrete Optimization},
  author={Korikov, Anton and Beck, J Christopher},
  booktitle={International Conference on Integration of Constraint Programming, Artificial Intelligence, and Operations Research},
  pages={18--34},
  year={2023},
  organization={Springer}
}

@inproceedings{korikov2021counterfactual,
  title={Counterfactual explanations via inverse constraint programming},
  author={Korikov, Anton and Beck, J Christopher},
  booktitle={27th International Conference on Principles and Practice of Constraint Programming (CP 2021)},
  pages={35--1},
  year={2021},
  organization={Schloss Dagstuhl--Leibniz-Zentrum f{\"u}r Informatik}
}

@article{kurtz2026counterfactual,
  title={Counterfactual explanations for linear optimization},
  author={Kurtz, Jannis and Birbil, {\c{S}} {\.I}lker and den Hertog, Dick},
  journal={European Journal of Operational Research},
  volume={329},
  number={1},
  pages={24--41},
  year={2026},
  publisher={Elsevier}
}

@article{mitrai2024takingpart1,
  title={Taking the human out of decomposition-based optimization via artificial intelligence, Part I: Learning when to decompose},
  author={Mitrai, Ilias and Daoutidis, Prodromos},
  journal={Computers \& Chemical Engineering},
  volume={186},
  pages={108688},
  year={2024},
  publisher={Elsevier}
}

\end{document}